\documentstyle[aps,prl,preprint,floats,epsfig]{revtex}

\def\NIMA{{\it Nucl. Instrum. Methods Phys. Res.}, Sec. A}

\def\PRL{{\it Phys. Rev. Lett.}}

\def\ZPC{{\it Z. Phys.} C}
\def\JMP{{\it Int. Jour. Mod. Phys.}}
\def\PRep{{\it Phys. Rep.}}
\def\etal{{\it et al.}}

\def\invfb{\rm fb^{-1}}

\def\eV{\rm eV}

\def\MeV{\rm MeV}
\def\GeV{\rm GeV}

\def\T{{\rm T}}
\def\CL{\rm CL}

\def\ra{\rightarrow}
\def\chisq{\chi^2}

\def\epl{e^+}
\def\emi{e^-}
\def\eplemi{\epl\emi}

\def\gg{\gamma \gamma}
\def\Ggg{\Gamma_{\gg}}
\def\pipl{\pi^+}
\def\pimi{\pi^-}

\def\piplpimi{\pipl\pimi}

\def\KS{K^{\circ}_S}
\def\KSKS{\KS\KS}

\def\fJ{f_J(2220)}

\def\qbar{\overline{q}}

\def\sbar{\overline{s}}

\def\K0{K^{\circ}}

\def\Kpl{K^+}
\def\Kmi{K^-}
\def\KplKmi{\Kpl\Kmi}

\def\B0{B^0}
\def\B0s{B^0_s}%
\def\B0d{B^0_d}%

\begin{document}

\preprint{\tighten\vbox{\hbox{\hfil CLNS 98/1560}
                        \hbox{\hfil CLEO 98-08}
}}

\title{\bf{FURTHER SEARCH FOR THE TWO-PHOTON PRODUCTION 
OF THE GLUEBALL CANDIDATE $\fJ$}}

\author{CLEO Collaboration}
\date{\today}

\maketitle
\tighten

\begin{abstract} 
The CLEOII detector at the Cornell $\eplemi$ storage ring CESR has been
used to search for the two-photon production of the $\fJ$ decaying into 
$\piplpimi$.
No evidence for a signal is found in data corresponding to an
integrated luminosity of $4.77\,\invfb$ and a $95\%\, \CL$ upper limit on 
$[\Ggg\, B_{\piplpimi}]_{\fJ}$ of $2.5\,\eV$ is set.
If this result is combined with the BES Collaboration's measurement of 
$\fJ \ra \piplpimi$ in radiative $J/\psi$ decay, 
a $95\%\, \CL$ lower limit on the stickiness of the $\fJ$ of $73$ 
is obtained.
If the recent CLEO result for $[\Ggg\, B_{\KSKS}]_{\fJ}$ is combined with
the present result, the stickiness of the $\fJ$ is found to be larger than 
$102$ at the $95\%\, \CL$.
These results for the stickiness (the ratio of the probabilities 
for two-gluon coupling and two-photon coupling) provide further support 
for a substantial neutral parton content in the $\fJ$.
\end{abstract}
\newpage

{
\renewcommand{\thefootnote}{\fnsymbol{footnote}}


\begin{center}
M.~S.~Alam,$^{1}$ S.~B.~Athar,$^{1}$ Z.~Ling,$^{1}$
A.~H.~Mahmood,$^{1}$ S.~Timm,$^{1}$ F.~Wappler,$^{1}$
A.~Anastassov,$^{2}$ J.~E.~Duboscq,$^{2}$ K.~K.~Gan,$^{2}$
T.~Hart,$^{2}$
% ...
\end{center}
 
\small
\begin{center}
$^{1}${State University of New York at Albany, Albany, New York 12222}\\
$^{2}${Ohio State University, Columbus, Ohio 43210}\\
% ...
\end{center}
 
\setcounter{footnote}{0}
}
\newpage


In lowest order, 
the two-photon width of a resonance is proportional to the fourth 
power of the constituent parton charges,
so a very small two-photon width is an indication of substantial neutral
parton content.
Within the framework of QCD, a small two-photon width implies that 
the resonance has substantial glueball content.
A quantitative measure of the glueball content of a resonance 
is the ratio of the probabilities for two-gluon coupling and two-photon 
coupling for which the resonance's two-gluon coupling is deduced from 
its production rate in radiative $J/\psi$ decay.

The $\fJ$ is a glueball candidate owing to its observation in radiative 
$J / \psi$ decay (a glue-rich environment)\cite{refMKIII,refBES},
its small two-photon width relative to its two-gluon 
width\cite{refARGUS,refCLEOglue}, 
its small total width\cite{refMKIII,refBES}, 
its similar branching fraction for non-strange and strange final 
states\cite{refBES}, and 
its proximity to the mass obtained in lattice 
calculations\cite{refMorningstar,refMichel} for a tensor glueball. 
CLEO has recently\cite{refCLEOglue} obtained a $95\%\, \CL$ upper 
limit on the product of the two-photon width and the $\KSKS$ branching 
fraction $[\Ggg\, B_{\KSKS}]_{\fJ}$ of $1.3\,\eV$ using the reaction
$\eplemi \ra \eplemi \fJ \ra \eplemi \KSKS$.
Earlier, the ARGUS Collaboration\cite{refARGUS} obtained a less 
restrictive limit based upon the $\KplKmi$ decay mode.
In the present paper we report on a search for the two-photon 
production of the $\fJ$ in the reaction
$\eplemi \ra \eplemi \fJ \ra \eplemi \piplpimi$.

The CLEOII detector\cite{refCLEOdet} is a general purpose detector 
operating at the Cornell Electron Storage Ring CESR\cite{refCESR}.
It provides charged particle tracking, 
precision electromagnetic calorimetry,
charged particle identification, and muon detection.
Charged particle detection over $95\%$ of the solid angle is provided 
by three concentric drift chambers in a magnetic field of $1.5\, \T$ 
giving a momentum resolution $\sigma_p / p = 0.5\%$ at 
$p = 1\, \GeV$.
The drift chambers are surrounded by a time of flight system and a 
CsI electromagnetic calorimeter.
A superconducting coil and muon detectors surround the calorimeter.
Two-prong events are recorded with three redundant triggers.
The results in this paper are based upon an integrated luminosity of
$4.77\,\invfb$ with CESR operating at a center-of-mass energy 
of approximately $10.6\, \GeV$.

The $\fJ$ is searched for in the two-photon reaction 
$\eplemi \ra \eplemi \fJ \ra \eplemi \piplpimi$
in the untagged mode in which the outgoing $\epl$ and $\emi$ are undetected.
Events are selected that have exactly two tracks of opposite charge whose
vector sum of momenta transverse to the beam has a magnitude 
less than $0.5\,\GeV$.
The total energy of the event is required to be less than $6.0\,\GeV$ and
the energy in the calorimeter not associated with either track must be less 
than $0.5\,\GeV$.

Two-photon produced final states of charged particle pairs are selected
(and backgrounds from Bhabha scattering, muon pair production, and
cosmic rays are suppressed) by requiring that the acolinearity of the 
two tracks is greater than $0.1$.
In addition, the acoplanarity is required to be less than $0.05$.
Here acolinearity is the deviation from colinearity 
in three dimensions while acoplanarity is the deviation from 
colinearity in the plane transverse to the beams.
These last two requirements are effective because the two-photon center-of-mass
generally moves rapidly and at a small angle with respect to the beams.

Events are vetoed if either track is identified as an electron or muon.
If $E / p$, the ratio of a track's energy deposition in the calorimeter and its
momentum measured in the drift chambers, is in the range 
$0.85 - 1.10$, the track is identified as an electron.
Muons are identified by the muon detectors.
Events must have satisfied at least one of the two-prong triggers.

The event simulation uses the BGMS\cite{refBGMS} formalism 
with the transverse-transverse term (appropriate for untagged two-photon
reactions) for the event generation and GEANT\cite{refGEANT} 
for the detector simulation. 
Photon form-factors based upon vector-meson dominance with a mass 
$m_V = 768.5\, \MeV$ are used. 
We take the spin of the $\fJ$ to be 2.
The detection efficiencies for helicity 0 and 2 are found to be
13.1\% and 26.9\% respectively.
We use a ratio\cite{refPoppe} of helicity 0 and helicity 2 of 1:6, 
giving an efficiency of 24.9\%.
When the mass $m_V$ in the photon form-factors is varied from 
$768.5\, \MeV$ to $\infty$ (corresponding to a form-factor equal to 1) 
the cross-section increased by 29.8\% while the efficiency dropped by 18.9\%
and their product increased by 5.5\%.
A $2.8\%$ systematic uncertainty is assigned to the product 
of the cross-section and efficiency.

The dominant source of systematic uncertainty is the trigger efficiency.
It is estimated to be $13\%$ from the observed variation 
of the event yield 
as a function of the azimuthal angle of each of the two tracks.
Data and simulation are compared to determine smaller systematic
uncertainties of 
$2\%$ per track from track reconstruction efficiency,
$3\%$ from the requirement on the energy deposition in the calorimeter, 
$3\%$ from the transverse momentum requirement, 
$2\%$ each from the acolinearity and acoplanarity requirements,
$5\%$ from the $E/p$ requirement, and
$4\%$ from the muon veto.
The total systematic uncertainty is the sum in quadrature 
of the above sources and is $15\%$.

A pion-pair invariant mass distribution is constructed using all events 
that pass the selection criteria and assuming that both particles are pions.
A plot of $m_{\pipl\pimi}$ in the mass region relevant for the
$\fJ$ is shown as the data points with statistical error bars in Fig.1.
\begin{figure}
\centering 
\leavevmode
\mbox{\epsfig{file=prlfig.eps,bbllx=70,bblly=280,bburx=550,bbury=730,
height=9.0cm,width=9.0cm,clip=}}
\caption{The $\pi^{+}\pi^{-}$ invariant mass distribution for the data
in the region of the $f_{J}$(2220).  
The hatched histogram is the expected signal shape with arbitrary 
normalization. 
The solid curve is the sum of a fit to the background and a signal 
corresponding to the 95\% CL upper limit on 
$\Gamma_{\gamma\gamma}B_{\pi^{+}\pi^{-}}$ of 2.5 eV.  
In the insert the two curves are the background fit 
with and without this level of signal added.}
\label{figMpipi}
\end{figure}
There is no evidence of an enhancement near the mass of the $\fJ$.
The mass distribution is fit with the sum of a signal and a background 
assuming that there is no interference between the two.
The signal shape is represented by a Breit-Wigner with a mean of 
$2231\,\MeV$\cite{refPDG97} and a width of $23\,\MeV$\cite{refPDG97} 
convolved with the detector resolution of $12\,\MeV$ and is shown 
as the hatched histogram in Fig.1.
The background is represented by a third order polynomial that is fit 
to the mass region $2000-2500\, \MeV$ excluding the region 
$2200-2268\,\MeV$.
The fit gives a signal of $-103 \pm 77$ events with a 
$\chisq = 35.6$ for 36 degrees of freedom.

An upper limit is obtained by only allowing for
a positive number of signal events, $N$.
Given that
$m_{\fJ} = 2231.1 \pm 2.5\,  \MeV$\cite{refPDG97} and 
$\Gamma_{\fJ} = 23 ^{+8}_{-7}\, \MeV$\cite{refPDG97},
likelihood functions for $N$ are obtained for a range of the resonance
mass and width, spanning $\pm 2.5 \sigma$ in each.  
These functions are then weighted with Gaussian probabilities 
for the mass and width to obtain a final likelihood
function $L_N$.
The product of the two photon partial width and charged di-pion branching
fraction, $\Ggg\, B_{\piplpimi}$, is 
given by the product of $N$ and $P$.  
Here $P$ is the partial width
used in the simulation divided by the product of luminosity, 
cross-section and efficiency; 
$P$ is assumed to be Gaussian distributed.
The likelihood function, $L_{\Gamma B}$
is then obtained 
by numerical integration in the two-dimensional space of $N$ and $P$.
From $L_{\Gamma B}$ a $95\%\, \CL$ upper limit 
of 2.5 eV for $\Ggg\, B_{\piplpimi}$ is obtained.
The solid line in the main portion of Fig.1 is the sum of the fit 
to the background and a signal that corresponds to this upper limit.
The mass region $2150-2310\, \MeV$ is shown enlarged in the inset
in Fig.1 with the two curves representing the background fit with and
without this level of signal added.

The upper limit can be specified without the assumption of a 1:6 
ratio for helicity 0 and 2 as
$(0.53 \Ggg^{2,0} + 1.08 \Ggg^{2,2}) B_{\piplpimi} < 2.5\,\eV$
at $95\%\, \CL$.
The superscripts indicate spin and helicity.
The ratio of the coefficients is equal to
the ratio of the efficiencies for helicity 0 and $\pm 2$ while the 
overall normalization is determined by the result given above.

The upper limit on $\Ggg\, B_{\piplpimi}$
can be interpreted in terms of the stickiness $S$\cite{refChanowitz}.
Stickiness is the ratio of the probabilities for two-gluon and 
two-photon coupling of a resonance, 
which in the present case can be written as ($f_J$ denotes $\fJ$):
\begin{eqnarray}
S_{f_{J}} = 
{|\langle f_{J}|gg \rangle|^2 \over |\langle f_{J}|\gg \rangle|^2} = 
C_{\ell} \left( {m_{f_{J}}\over k_{\gamma }} \right)^{2\ell+1} 
{\Gamma_{J/\psi} B(J/\psi \ra \gamma f_{J}) B(f_{J} \ra \pi^{+}\pi^{-}) \over
\Gamma (f_{J} \ra \gg) B(f_{J} \ra \piplpimi)}
\end{eqnarray}
The parameter $k_{\gamma}$ is the energy of the photon produced in the 
radiative $J/\psi$ decay as calculated in the $J/\psi$ rest frame, 
and $\Gamma_{J/\psi}$ is the total width of the $J/\psi$.
The factor with $2\ell + 1$ in the exponent 
removes the trivial phase space
dependence of the stickiness upon the $f_{J}$ mass.
The quantum number $\ell$ is the relative angular momentum between 
the two gluons or photons, with 
$\ell = 0$ for $J = 2$.
$C_0 = 20.5$ is a normalization factor 
chosen such that the stickiness is normalized to unity for
the $f_2(1270)$.
The BES result \cite{refBES} and $J / \psi$ properties
from the Particle Data Group\cite{refPDG97} are 
combined with our result to obtain a likelihood distribution for
the stickiness of the $f_{J}$ via a Monte Carlo technique.
In this procedure the $L_{\Gamma B}$ obtained previously was used
and all other uncertainties were taken to be Gaussian distributed.  
A lower limit of $S_{f_{J}} > 73$ is found at $95\%\, \CL$.

This lower limit and the one obtained in the $\KSKS$ 
channel\cite{refCLEOglue} can be merged, 
again using a Monte Carlo procedure, 
to obtain a combined lower limit\cite{refComment} 
on the stickiness of $S_{f_{J}} > 102$, also at $95\%\, \CL$.
This result can be compared with the stickiness of the $f_2'(1525)$,
a resonance thought to be predominantly an $s \sbar$ bound state.
Using the properties of the $f_2'(1525)$ from the 
Particle Data Group\cite{refPDG97} a stickiness 
$S_{f_2'} = 14.7\pm 3.9$ is found, considerably smaller than the 
lower limit $S_{f_{J}} > 102$.
A linear superposition of $|q \qbar>$ states can be constructed such that
the two-photon width is negligible; the coefficients would have to take
on very specific values so this possibility is considered unlikely.
The large lower limits on the stickiness of the $\fJ$ are therefore
an indication of substantial neutral parton or glueball content.

In this Letter a restrictive $95\%\, \CL$ upper limit 
$[\Ggg B_{\piplpimi}]_{\fJ} < 2.5\,\eV$ is presented.
Using the BES Collaboration's result for $\fJ \ra \piplpimi$ 
in radiative $J/\psi$ decay,
this upper limit leads to a lower limit on its stickiness
$S_{\fJ} > 73$ at $95\%\, \CL$.
When these results are combined with an earlier 
CLEO result\cite{refCLEOglue}, 
a lower limit on the stickiness of $102$ at $95\%\, \CL$ is obtained.
This large value is difficult to understand if the valence partons 
of the $\fJ$ are quarks and antiquarks only;
therefore, the $\fJ$ is likely to have a substantial neutral parton
or glueball content.

We gratefully acknowledge the effort of the CESR staff in providing us with
excellent luminosity and running conditions.
This work was supported by 
the National Science Foundation,
the U.S. Department of Energy,
Research Corporation,
the Natural Sciences and Engineering Research Council of Canada, 
the A.P. Sloan Foundation, 
the Swiss National Science Foundation, 
and the Alexander von Humboldt Stiftung.


\begin{thebibliography}{99}
\bibitem{refComment}In this report and in Ref. \cite{refCLEOglue} the 
uncertainty on $C_0$ is not included because the stickiness could have
been normalized to any $J = 2$ resonance that is thought to 
have light quarks as valence partons, 
leading to different values for the stickiness.
If that uncertainty is included, the $95\%\, \CL$ lower limits on the 
stickiness of the $\fJ$ are $69$ and $67$ for the $\KSKS$ and $\piplpimi$
channels respectively (instead of $76$ and $73$) while the combined 
$95\%\, \CL$ lower limit is $94$ (instead of $102$).
\end{thebibliography}
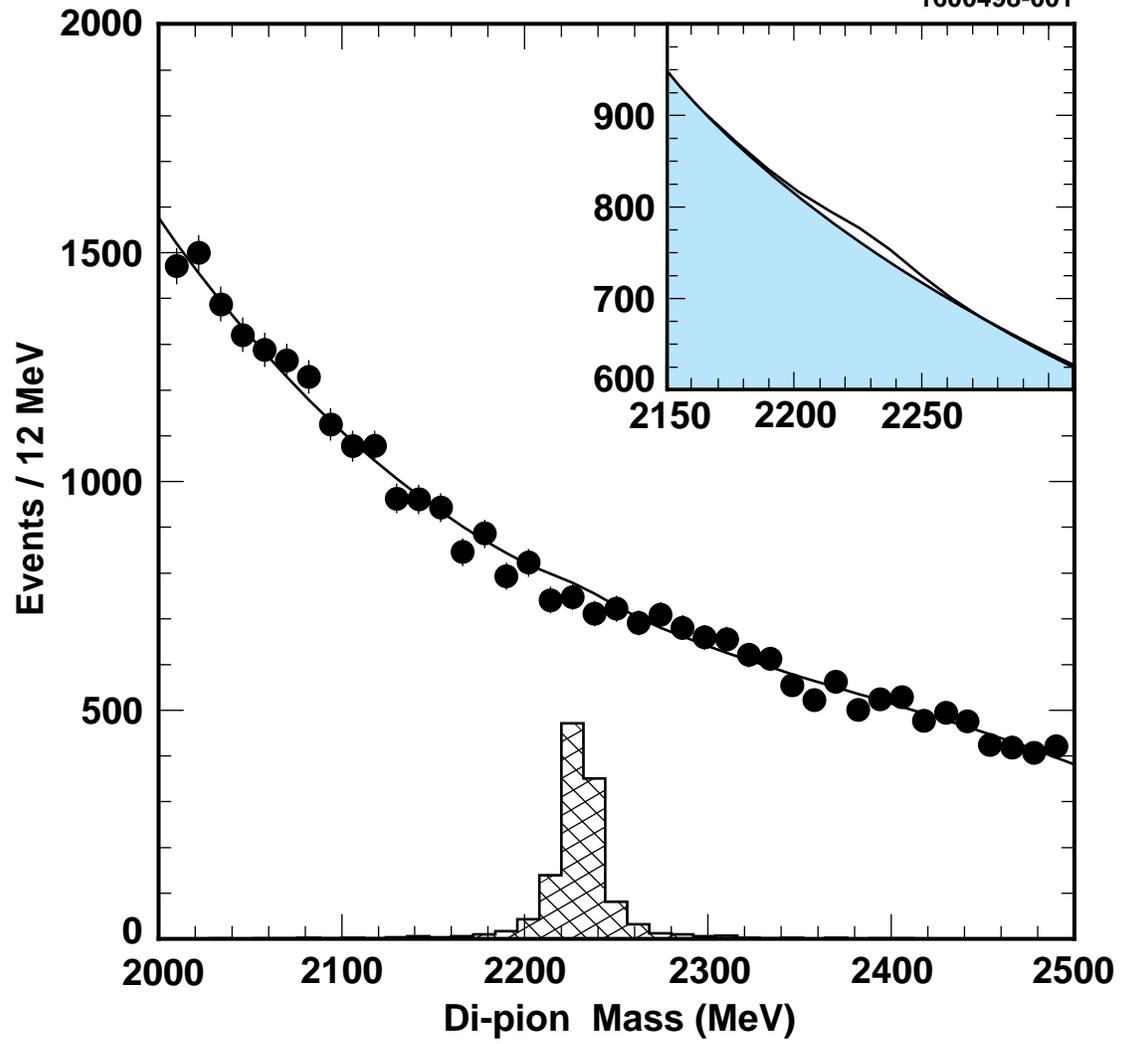
\end{document}